\documentclass[preprint,3p,number]{elsarticle}
\usepackage{xcolor}
\usepackage{subfigure}
\usepackage{graphicx}
\usepackage{array}
\PassOptionsToPackage{hyphens}{url}
\usepackage{hyperref}

\title{Design of high-strength, radiopure copper-chromium alloys for rare-event searches assisted by computational thermodynamics}
\author{D. Spathara$^a$}
\ead{d.spathara@bham.ac.uk}
\affiliation{organization={School of Physics and Astronomy, University of Birmingham},
postcode={B15 2TT},
state={Birmingham},
country={United Kingdom}}

\begin{document}
\begin{abstract}
Direct Dark Matter detection and studies on the nature of neutrinos demand detector systems with extremely low background levels, including from radioactivity. Additive-free, electroformed copper, in addition to a set of advantages, exhibits exceptional radiopurity, making it the material of choice for mechanical and structural components for rare-event searches experiments. To satisfy the increasing demand for materials with superior mechanical strength, the development of copper-chromium alloys is pursued. Early investigations explored the synthesis of these alloys by electrodeposition and thermal processing. A materials-design approach is proposed to optimize the fabrication and thermal processing stages of manufacturing. 
It is assisted by materials modeling tools based on the thermodynamic and kinetic properties of alloy compositions, which enables faster development of novel materials by predicting properties and materials performance. This approach is demonstrated by comparing simulations with previously reported experimental investigations and proposing improved thermal processing.
\end{abstract}


\maketitle

\section{Introduction}
Immense progress has been achieved in understanding the Cosmos over more than 45 orders of magnitude in length. However, fundamental questions remain about the nature of dark matter (DM) and neutrinos. The existence of DM, corresponding to 84.5\% of matter in the Universe, is established via astrophysical observations and precise cosmological measurements~\cite{Planck:2018vyg}. Several search approaches are used, with “direct detection” being one of the most prominent: to detect DM from the Milky Way halo through its coherent elastic scattering off a nucleus~\cite{Billard:2021uyg}. At the same time, clarification of whether neutrinos are Majorana or Dirac particles will have profound implications and may lead to the discovery of new physics phenomena. A most sensitive probe is the search for neutrinoless double $\beta$-decay, a process only possible for Majorana neutrinos~\cite{Dolinski:2019nrj}.

Copper is the material of choice for structural detector components used for rare-event searches. It is commercially available at low cost and high purity, has excellent thermal and electrical conductivity, and has no long-lived radioactive isotopes. However, it may be contaminated during manufacturing and can be activated by fast neutrons from cosmic rays. To meet the unique radiopurity requirements, attention is focused on electroformed copper (EFCu), which can be plated at an electric potential that suppresses depositions of Th and U contaminants. This novel technique leads to extremely low contamination below 10$^{-14}$ grams of $^{232}$Th and $^{238}$U per Cu gram~\cite{MAJORANA:2016lsk, NEWS-G:2020fhm}.  Rare-event search experiments pioneer large-scale, additive-free Cu electroformation, e.g. the completed Majorana demonstrator~\cite{MAJORANA:2016lsk} and the on-going ECuME project~\cite{knights2023status}. In addition, to support the construction of experiments, underground laboratories provide manufacturing of EFCu as a service~\cite{Borjabad:2018wda}. 

Despite its advantages, Cu is highly ductile and of low strength, which limits its use for moving mechanical, high-pressure, and load-bearing parts. Alloying Cu with other elements, e.g. Be, Cr, Zr, Ti, Ni, Si can enhance its mechanical properties~\cite{zhang2019review, yang2023recent, HUANG2021102378}. This is of interest for a wide range of applications, including transportation, electronics, and electric circuits~\cite{yang2023recent, peng2005property, yuan2017microstructure, xu2018effect}. However, during manufacturing, the alloy needs to undergo a thermal strengthening process which may affect its electrical properties. Thus, a trade-off between strength and electrical conductivity needs to be made, in addition to the balance between strength and radiopurity, relevant for rare-event searches. 

Since Cr can be plated on the Cu surface and vice versa, copper-chromium (CuCr) alloys  have the potential to fulfill the enhanced strength and radiopurity requirements, and are of particular interest for rare-event searches. For example, the DarkSPHERE experiment~\cite{NEWS-G:2023qwh,Knights:2025ogz}, proposes to use a large-scale high-pressure fully electroformed underground spherical proportional counter to probe uncharted territory in the search for sub-GeV particle DM. An alloy of enhanced strength would enable detector operation at higher pressure and improved physics reach. Another example is the nEXO experiment, a neutrinoless-double $\beta$-decay experiment aiming at clarifying the nature of neutrinos~\cite{nEXO:2021ujk}, which requires a pressure vessel capable of containing liquid xenon. This could also be realized through a high-strength, radiopure alloy. 

We aim to use materials modeling tools, based on computational thermodynamics, to design the manufacturing of CuCr alloys, including electroformation and thermal processing, with optimal parameters. In this work, as a first step, simulations are compared with results of published experimental investigations, and improved processes are suggested. To the best of author's knowledge, this is the first time computational thermodynamics has been applied in the design of materials for fundamental particle physics research. Further work towards synthesis of application specific Cu-based alloys and their impact to rare-event search experiments is reported in Ref.~\cite{Spathara:2025hrp}.

\section{Experimental investigations of CuCr alloys}
The advantage of Cu plating at much lower voltage, i.e. approximately 0.3~V, is that it suppresses deposits of Th or U, which are present in the electrolytic bath in the form of ions. Cr, on the other hand, needs to be plated at higher voltage, e.g., in the vicinity of 0.5~V, with the risk of increasing Th and U contamination. However, CuCr alloys are projected to have radiopurity at levels similar to those of EFCu~\cite{Suriano:2018nrb}.

Small additions of Cr to Cu lead to a substantial increase in the strength of the CuCr alloy~\cite{HUANG2021102378}. This has also been demonstrated in initial experimental investigations aiming at manufacturing electroformed CuCr alloys~\cite{Suriano:2018nrb, Vitale:2021xrm}, while maintaining a high radiopurity. Local measurements of CuCr at 0.585~wt\% Cr indicated that it can improve the hardness of EFCu by 70\%~\cite{Suriano:2018nrb, osti_1039850}. A single point measurement of CuCr material with a maximum composition of approximately 0.4~wt\% Cr points to maximum hardness above 140~HV (Vickers hardness); i.e. 100\% harder than EFCu found in the same study~\cite{Vitale:2021xrm}.

However, co-electrodeposition of Cu and Cr is particularly challenging~\cite{Vitale:2021xrm}. Both electroformed Cu and Cr electroplating need to be fabricated in an additive-free, water-based solution to avoid further contamination with radioactive backgrounds. This poses difficulties to control the concentration of different species in the electrodeposits and the surface roughness and growth. Initial investigations into the synthesis of radiopure CuCr alloys reported electrodeposition of Cr and Cu layers, electrodeposited at different times using separate solution baths~\cite{Suriano:2018nrb, Vitale:2021xrm}, followed by solution heat treatment at 1025$^{\circ}$C for 24~h and aging at 500$^{\circ}$C for 12~h~\cite{Vitale:2021xrm}. 

Solution heat treatment is required so that a homogenized composition can be achieved in a single phase, before quenching - i.e. cooling at a very fast rate. Pure Cr and Cu are organized thermodynamically in different crystal structures. Cr is found in the body centred cubic (bcc) and Cu in the faced centred cubic (fcc) crystal structure. At a temperature high enough for a single phase to be thermodynamically stable, after phase transformation occurs, Cu and Cr atoms are organized in an fcc crystal structure. Cr atoms diffuse from the bcc structure until the whole volume of the bcc phase is transformed into a single fcc matrix phase, leading to a supersaturated solid solution after quenching. At this point, the solution heat treatment has been completed. The misfit between the different sizes of the Cu and Cr atoms occupying the sites of the fcc lattice leads to lattice distortion and matrix strengthening by hindering dislocations movement. In addition to the difference in the size of atoms of the solute in the matrix, the concentration of the solute atoms is also important~\cite{zhang2019review, yang2023recent}.

Aging aims at precipitation strengthening and follows solid solution strengthening. 
Starting from a fully homogenized alloy composition and a single-phase microstructure, the controlled growth of a secondary phase, i.e. bcc, precipitates into the Cu-rich matrix during aging. The two-phase microstructure is thermodynamically stable in a temperature range lower than that at which homogenization can be achieved. Cr diffuses from the matrix to form spherical precipitates. The incoherence between the two phases and the size of the spherical precipitates are the main reasons for mechanical strengthening. During precipitation strengthening, the formation of precipitates results in a decrease in the concentrations of solute atoms in the matrix, which reduces the impact on electrical conductivity~\cite{yang2023recent}.

Aging is an evolving process, where the secondary phase starts as an embryo and gradually reaches the size of the spherical precipitates at a critical radius, at which the size is ideally very small, and the volume fraction of the precipitates is the maximum possible for the specific alloy composition. 
Solid solution strengthening, previously achieved through solution heat treatment for alloy homogenization, is not as pronounced at this stage, where heat treatment is taking place at a much lower temperature than the homogenization stage. Because Cr atoms will diffuse from the matrix to the precipitating phase, the fcc lattice distortion decreases and so will the yield strength due to solid solution strengthening. 
By extending aging for longer periods, larger precipitates evolve by a simultaneous dissolution of smaller precipitates at the coarsening stage, leading to a lower yield strength increase due to precipitation strengthening.

In the process of improving mechanical properties of CuCr alloys for low background application, two strengthening mechanisms are crucial, namely solid solution strengthening and precipitation strengthening. An example from industrial application, demonstrated that Cu-0.5Cr (in wt\%) alloy exhibits hardness above 160~HV, tensile strength of 499~MPa, and electrical conductivity of 74.8~\%IACS (International Annealed Copper Standard), after aging for 1 hour at 450$^{\circ}$C~\cite{HUANG2021102378}. 
In this investigation, the alloy was fabricated from casting, solution heat treated for 3 hours at 1050$^{\circ}$C, and mechanically strengthened by cold rolling. Although casting and mechanical deformation compromise radiopurity, and thus cannot be used in the application at hand, the modeling aspects for thermal processing, especially for the aging stage, are relevant for this investigation.

In order to optimize the parameters of temperature and time for both solution heat treatment and aging stages, a thermodynamic and kinetic description of the Cu-Cr system is required. For example, the solubility of Cr in Cu is limited and only a narrow single-phase field of less than 1~wt\% can be found, in the range of 1020$^{\circ}$C – 1080$^{\circ}$C pointing to temperatures that could be selected for the solution heat treatment stage~\cite{zeng1995thermodynamic, jacob2000thermodynamic, turchanin2006phase}. After quenching, the combination of temperature and duration of time needs to be carefully selected to optimize the precipitation strengthening stage with aging. Since the size of precipitates and their distribution in the matrix determine the mechanical enhancement of Cu-based alloys~\cite{yang2023recent, HUANG2021102378, tang2018kinetic}, the optimized parameters in the design of the thermal process for specific alloy compositions can be determined using computational thermodynamics~\cite{tang2018kinetic, national2008integrated,lu2014computational,LUO20156,de2019new, li2021calphad}.

\section{Modeling Methods}
The modeling using computational thermodynamics aims to determine the optimal time and temperature for the homogenization and aging stages. These tools are constructed within the CALPHAD framework (CALculation of PHAse Diagrams)~\cite{lukas2007computational, SPENCER20081, kattner2016calphad}. 
DICTRA and TC-PRISMA, presented in the following, are kinetic simulation modules, in the Thermo-calc Software package. In this work, Thermo-Calc  version 2024a is used, and the CALPHAD-type databases Thermo-calc Software TCHEA6 and MOBHEA3 High Entropy alloys database~\cite{TCdatabase1, TCdatabase6}, for the description of the thermodynamic and kinetic properties of the Cu-Cr system. 

\subsection{Solution heat treatment (homogenization)}
For the solution heat treatment stage, to achieve homogenization, DICTRA 1D simulations are employed~\cite{andersson2002thermo}. 
These solve the diffusion equations at the interface of two regions of different composition during phase transformation. This is done at an isothermal, i.e. a single temperature for a specific duration of time, in a volume-fixed frame of reference~\cite{andersson1992models,spathara2018study}.

The setup consists of two regions, separated by a planar boundary, with initial concentrations of Cr and Cu and phases of bcc and fcc, respectively, at each region. For each of them, the single-phase model~\cite{andersson2002thermo} is used. Because Cr and Cu are found in different phases, the moving boundary model~\cite{andersson2002thermo} is applied, where the phase transformation occurs at the interface of the two regions. Assuming a local two-phase equilibrium and mass conservation at the interface, the simultaneous solution of the equations yields the interfacial concentrations and the growth rates of each species.  
At the end of each simulation, the concentration profiles of each species are obtained. 

In the diffusion equations, the expression of the flux $J_{i}$ of each component (e.g. Cr) can be expanded in terms of concentration gradients $\vartheta{c_{j}}$ in direction $z$:
\[
J_{i} = - \sum^{n}_{k=1} D_{ij}\frac{\vartheta{c_{j}}}{\vartheta{z}}
\]
where $D_{ij}$ is the diffusion coefficient of component $i$ (Cr or Cu) with respect to the concentration gradient of component $j$ (Cu or Cr, respectively).

The flux-balance equation ensuring mass conservation of each component determines the migration rate at the phase interface:
\[
\nu^{bcc/fcc} (c^{bcc}_{Cr} - c^{fcc}_{Cr}) = J^{bcc}_{Cr} - J^{fcc}_{Cr}
\]
where $\nu^{bcc/fcc}$ is the interface velocity, $c^{bcc}_{Cr}$ and $c^{fcc}_{Cr}$ are the Cr concentrations in the bcc and fcc phases, respectively, and $J^{bcc}_{Cr}$ and $J^{fcc}_{Cr}$ the fluxes of Cr on each side of the interface. Integration in time includes initial calculation of boundary conditions at the interface, by combining the moving boundary and single-phase models.

\subsection{Aging (precipitation strengthening)}
The nucleation and growth of precipitates during aging are simulated using TC-PRISMA~\cite{tang2018kinetic}, according to the Langer-Schwartz theory~\cite{langer1980kinetics}. In addition, the Kampmann Wagner numerical method is adopted~\cite{wagner2001homogeneous}. In this way, the nucleation and growth of the precipitating secondary phase are considered simultaneously. At the end of each simulation, the mean radius, the volume fraction, and the particle size distribution of the precipitates are obtained.

According to the classical nucleation theory, on which the Langer-Schwartz theory is based,  the nucleation rate $J_{s}$ for a steady state is given by the following:
\[
\ J_{s} = Z \beta^{*} N \exp{({-\frac{\Delta G^{*}}{kT}})}
\]
where $\Delta G^{*}$ is the nucleation barrier, i.e. the energy cost associated with forming a stable nucleus of the new phase, $Z$ is the Zeldovich factor, $\beta^{*}$ the atomic attachment rate, and $N$ the number of nucleation sites. 
The nucleation barrier is given by:

\begin{equation} \label{eq5}
\ \Delta G^{*} = \frac{16 \pi \sigma^{3} V^{2}_{m}}{3\Delta G^{2}_{m}}
\end{equation}
where $\Delta G_{m}$ is the driving force of the multicomponent system, i.e. the thermodynamic potential that favors the formation of the new phase, $V_{m}$ is the molar volume of the precipitating phase, and $\sigma$ is the interfacial energy, where the value of 0.7~$J/m^2$ is adopted for Cu-0.5Cr~(in~wt\%) following Ref.~\cite{HUANG2021102378}.  

The time-dependent nucleation rate for a non-steady state $J(t)$ is related to $J_s$ as: 
\[
\ J(t) = J_{s} \exp{({-\frac{-\tau}{t}})}
\]
where $\tau$ is the incubation time. The parameters appearing in these equations are determined, along with many other interesting parameters, from the model introduced in Ref.~\cite{chen2008analytical}.

\subsection{Solution and precipitation strengthening}
The yield strength due to solid solution strengthening, $\Delta\sigma_{ss}$,  is determined by the Fleischer equation~\cite{gottstein2013physikalische, freudenberger2010non}:
\begin{equation} \label{eq6}
\ \Delta\sigma_{ss} = G (|\delta|+\frac{1}{20}|\eta|)^{3/2}~\sqrt{\frac{\chi_{\rm Cr}}{3}}
\end{equation}
where $\chi_{\rm Cr}$ is the mole fraction of the solute species (Cr, in at\%) in the matrix, 
$G$ is the shear modulus of the matrix (for Cu), and $\delta$ and $\eta$ are factors associated with the lattice misfit and the variation due to the shear modulus, respectively.

The yield strength due to precipitation strengthening, $\Delta\sigma_{os}$, is given by the Orowan mechanism~\cite{martin2012precipitation, HUANG2021102378}:
\begin{equation} \label{eq9}
 \Delta\sigma_{os}= \frac{0.81~G~b}{2\pi(1-\nu)^{1/2}} \frac{ln(2r/b)}{(\lambda - 2r)}
\end{equation}
where $b$ the Burgers vector for the Cu matrix, $\nu$ the Poisson's ratio for Cu, $r$ the mean particle radius, $f$ the precipitate volume fraction, and $\bf{\lambda}$ the effective interparticle spacing~\cite{HUANG2021102378}: 
\begin{equation} \label{eq10}
\ \lambda = \Bigl(\sqrt{\frac{2\pi}{3f}} - 1.63\Bigl)r
\end{equation}
Table~\ref{tab1} lists the inputs used to determine the yield strengthening.
\begin{table}[h!]
    \centering
    \caption{Parameters for Yield strength evolution}
    \footnotesize\setlength{\extrarowheight}{2pt}
    \begin{tabular}{l l l}\hline
    \underline{Parameter}&\underline{Value}&\underline{Ref.}\\
    shear modulus $G(Cu)$&46~GPa&\cite{gottstein2013physikalische, HUANG2021102378}\\
    factor $\delta$ &0.0435& \cite{HUANG2021102378}\\
    factor $\eta$ &0.1492& \cite{HUANG2021102378}\\
    Burgers factor b&0.2556~nm&\cite{freudenberger2010non, HUANG2021102378}\\
    Poisson's ratio $\nu$&0.34&\cite{freund2004thin, HUANG2021102378}\\
    Cr atomic frac. $\chi_{Cr}$& calculated (at\%)&this work \\
    mean radius of precip. $r$&calculated (nm)&this work\\
    precipitate volume fraction $f$ &calculated&this work\\\hline
    \end{tabular}
    \label{tab1}
\end{table}

\section{Results and discussion}
The results obtained from simulations of both solution heat treatment and aging are presented and compared to the literature. 
\subsection{Solution heat treatment}
Aiming at homogenization, simulations are performed for 5 isothermals, at 950$^\circ$C, 1000$^\circ$C, 1050$^\circ$C and 1060$^\circ$C, for 24 hours each. Simulations for the 1025$^\circ$C are also extended to a 72-hour isothermal. The three lower temperatures were selected to match those used in Ref.~\cite{Vitale:2021xrm}. 
The initial configuration of the Cu and Cr layers is also chosen to match those used in Ref.~\cite{Vitale:2021xrm}:
a) 1~$\mu m$ of Cr, and
b) 10~$\mu m$ of Cr, 
in contact with 400~$\mu m$ of Cu substrate.

\begin{figure}
    \centering
    \includegraphics[width=0.45\linewidth]{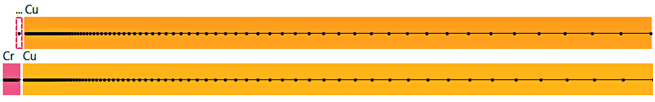}
    \caption{Schematic representation of grid points in the DICTRA setup for 400$\mu m$~Cu in contact with Cr thickness layer of: 1~$\mu m$ (top) and 10~$\mu m$ (bottom).     
    \label{fig:0}}
\vspace{-0.3cm}
\end{figure}

\begin{figure*}[h]
    \subfigure[\label{fig:1a}]{\includegraphics[width=0.48\linewidth]{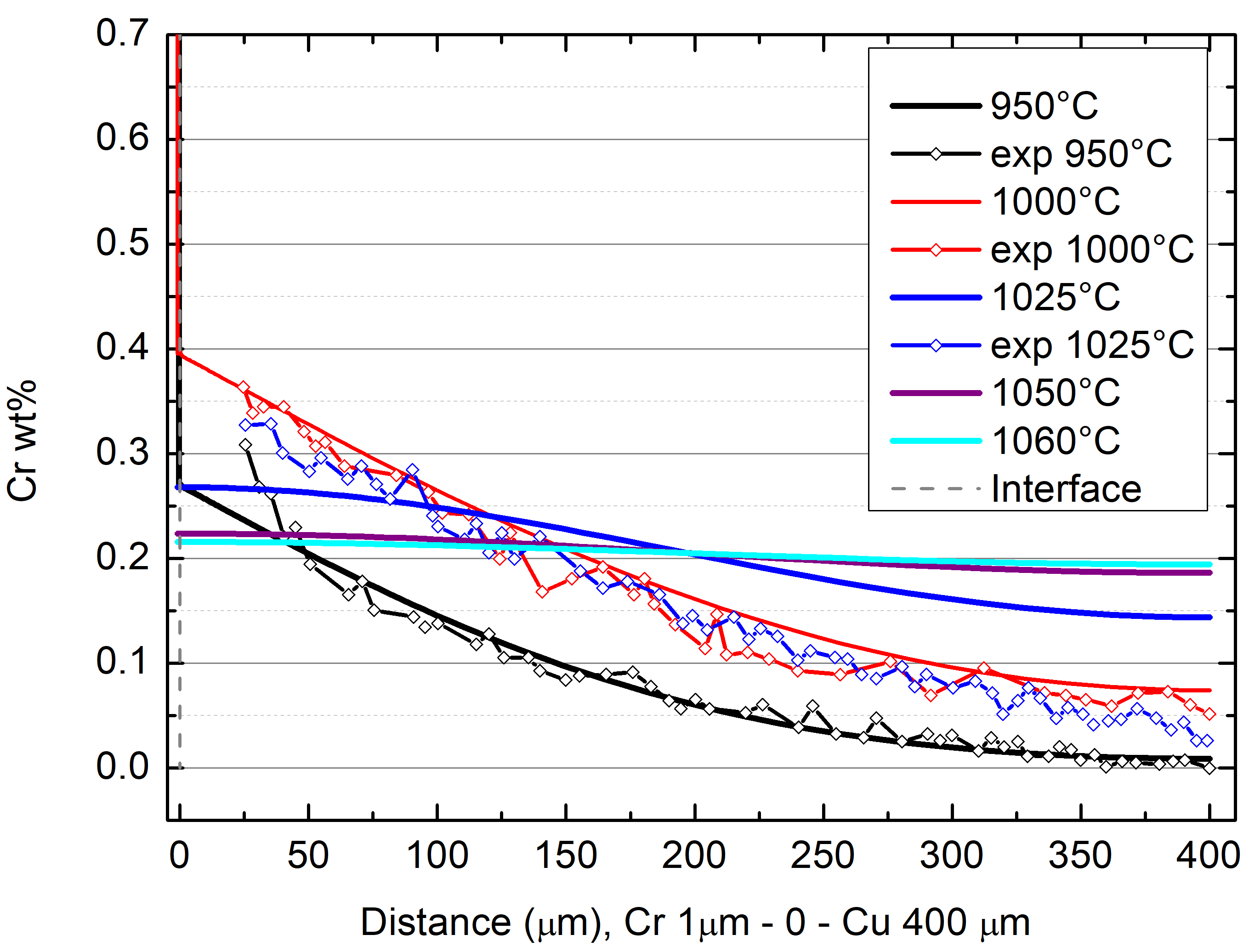}}
    \subfigure[\label{fig:1b}]{\includegraphics[width=0.48\linewidth]{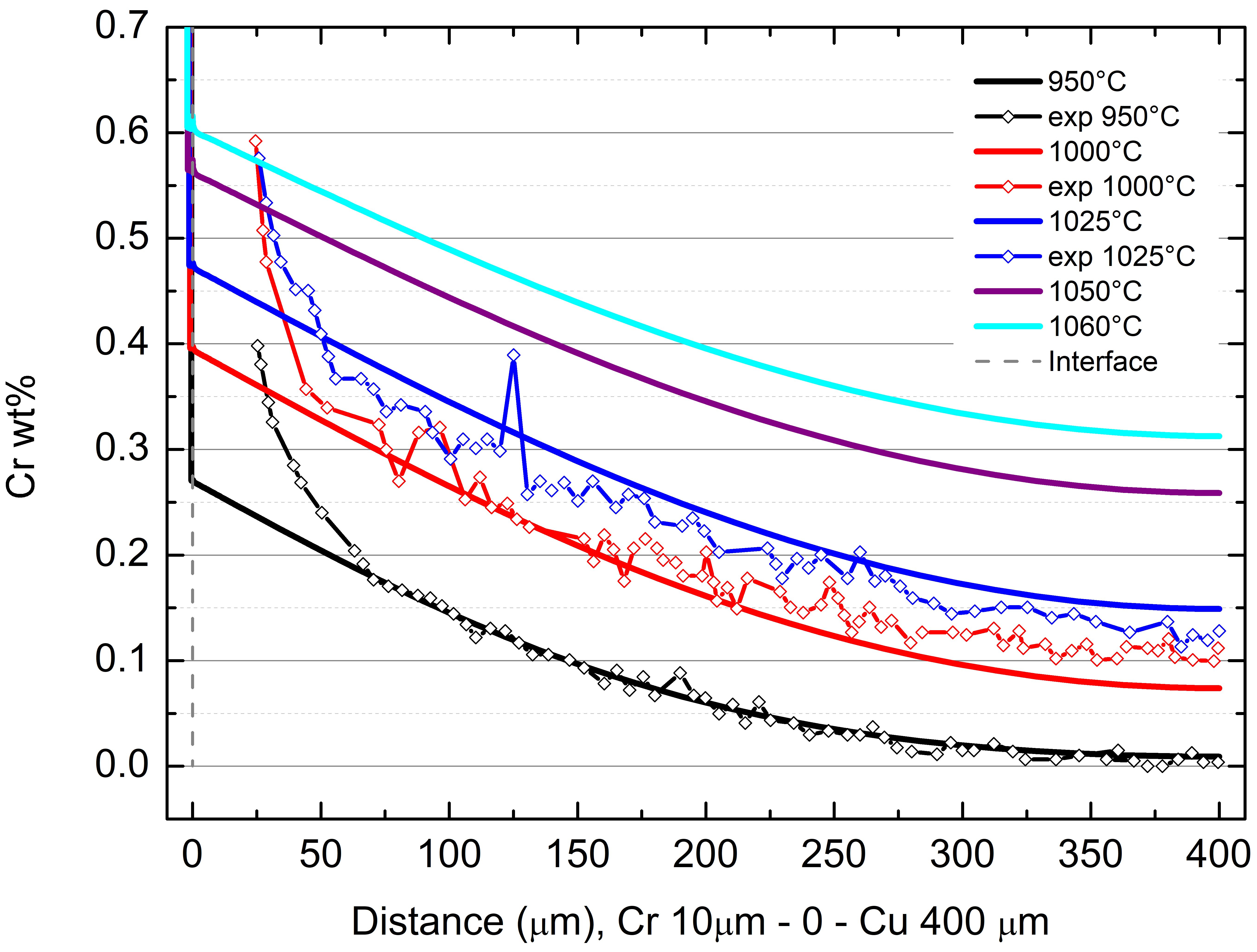}}
    \caption{Cr concentration profiles of 400 $\mu m$ Cu in contact with \subref{fig:1a} 1 $\mu m$ Cr and \subref{fig:1b} 10 $\mu m$ Cr, using 1D DICTRA simulations for 24 hours. Experimental data corresponding to solution heat treatment, with aging at 500$^\circ$C is included~\cite{Vitale:2021xrm}.\label{fig:1}}
    \vspace{-0.3cm}
\end{figure*}
The grid is set-up in a planar geometry.
For the Cu region 400 geometric grid points of initial 100\%~Cu concentration are selected such that the density is highest toward the interface with the Cr region. For the Cr region, in the 1~$~\mu m$ layer case, 
50 linear grid points of 100\%~Cr initial concentration are selected, while in the 10~$\mu m~Cr$ layer 100 grid points are selected. The grid configuration is shown in Fig.~\ref{fig:0}.

\begin{figure}
    \centering
    \includegraphics[width=0.45\linewidth]{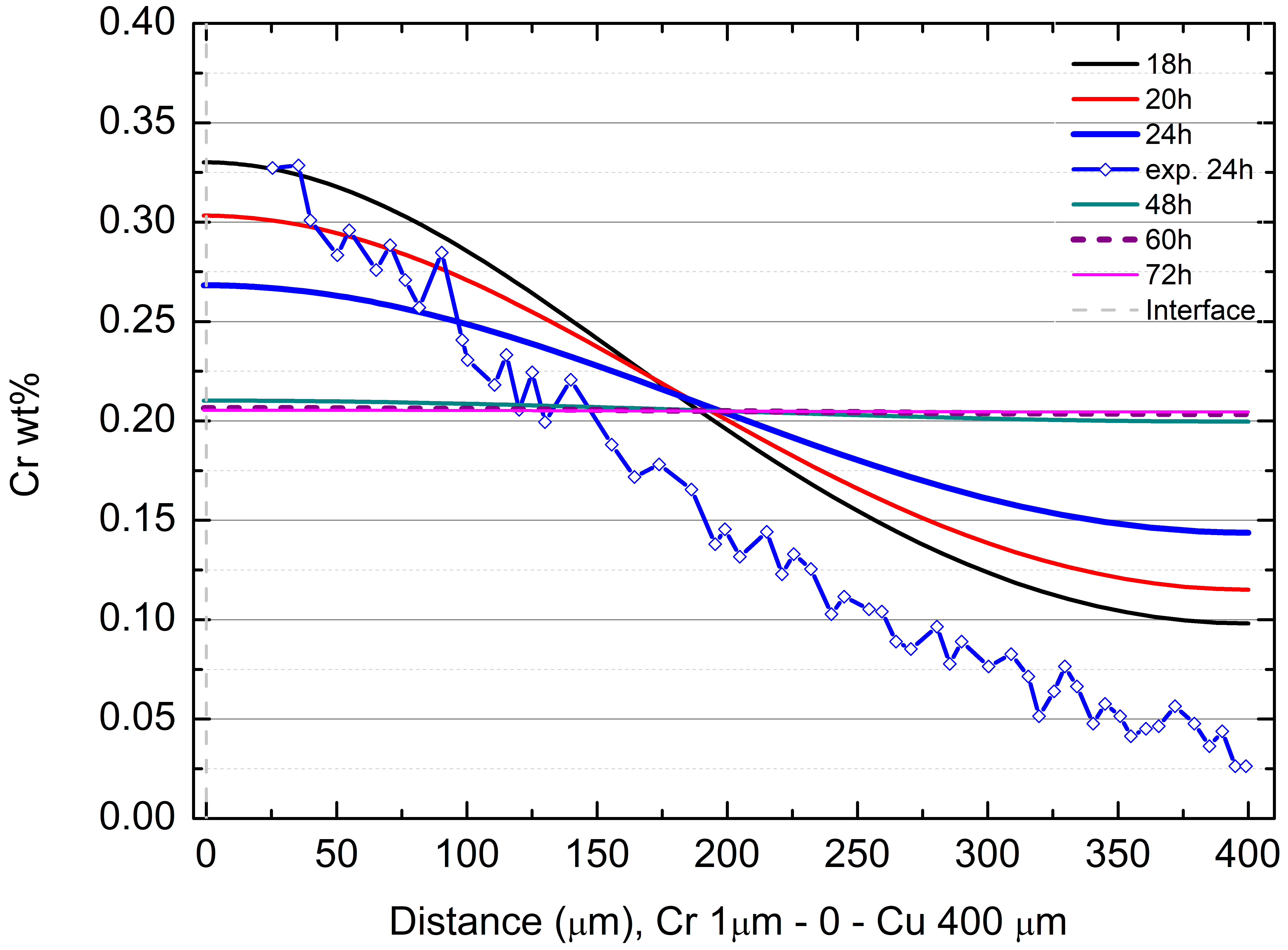}
    \caption{Cr concentration profile of 400 $\mu m$ Cu in contact with 1 $\mu m$ Cr at 1025$^\circ$C. Experimental data for 24 hours isothermal, with aging at 500$^\circ$C is included~\cite{Vitale:2021xrm}.\label{fig:72h}}
    \vspace{-0.3cm}
\end{figure}

Figure~\ref{fig:1} shows the Cr concentration profile obtained after 1D simulations using DICTRA for the two cases.
Experimental measurements obtained from heat-treated and aged samples for three out of five temperatures are included from Vitale et al.~\cite{Vitale:2021xrm}. 

According to simulations, for the 1~$\mu m$~Cr thickness (Fig.~\ref{fig:1a}), a single phase can be achieved after 24 hours at 1025$^\circ$C, 1050$^\circ$C and 1060$^\circ$C. A homogeneous alloy composition of about 0.2~wt\% Cr is obtained from simulations at 1050$^\circ$C and 1060$^\circ$C. In the case of 1025$^\circ$C, a longer duration is required to achieve homogenization, namely 72~h, as shown in Fig.~\ref{fig:72h}. Thermodynamic stability is achieved faster at higher temperatures, as shown from the simulation results in Fig.~\ref{fig:1a} and Fig.~\ref{fig:72h}. 

The different solubility of Cr in Cu at different temperatures, according to the Cu-Cr phase diagram, becomes apparent in Fig.~\ref{fig:1b}, where the thickness of the Cr layer exceeds the amount of Cr that can be incorporated in a CuCr alloy during the 24~h isothermals.

In the case of 1 $\mu m$ Cr in contact with 400~$\mu m$ Cu, simulations of the solution heat treatment stage are found to be in agreement with the experimental data of the heat-treated and aged samples of Ref.~\cite{Vitale:2021xrm}, in particular for the 950$^\circ$C and 1000$^\circ$C 24-hour isothermals. It is noted that in Ref.~\cite{Vitale:2021xrm} Cr concentration profiles in the Cu region are reported at a distance larger than 25~$\mu m$ from the interface. Moreover, although the measurements in Ref.~\cite{Vitale:2021xrm} correspond to samples that are also aged, the observed trend is reproduced by the simulations, in particular for the lowest temperatures. 

The zigzag Cr concentration profiles reported in Ref.~\cite{Vitale:2021xrm} reveal the presence of Cr-rich precipitates, especially at a higher temperature of 1025$^\circ$C. This is a result of the second heat treatment at 500$^\circ$C for 12~h aiming at aging. Since measurements in Ref.~\cite{Vitale:2021xrm} indicate that homogenization has not been achieved after thermal processing, the more pronounced zigzag trend closer to the Cr/Cu interface can be explained by the higher Cr concentration in the Cu region. This is accentuated in the case where the solution heat treatment takes place at higher temperatures, which favour higher Cr concentration in the Cu region. As observed, for the solution heat treatment at 950$^\circ$C the zigzag pattern is less pronounced.

Because the Cr bcc phase is transformed into Cu fcc in the interface during solution heat treatment, the sharp interface in simulations moves toward the Cr region until homogenization is achieved. Vitale et al.~\cite{Vitale:2021xrm} reported higher Cr concentrations in the Cu region. The observed discrepancy between the simulations and measurements closer to the Cr/Cu interface, especially in the case of 10~$\mu m$-thick Cr layer, could be explained by the roughness of the substrate prior to Cr electrodeposition. Given that a single phase cannot be achieved in the case of the thicker Cr region (10~$\mu m$), this results in a more pronounced discrepancy closer to the interface of the two phases.

From the presented comparisons, we deduce that the optimal solution heat treatment is the 24~h at 1050$^\circ$C, rather than the 1025$^\circ$C employed in Ref.~\cite{Vitale:2021xrm}. In addition, the 1 $\mu m$ Cr layer in contact with 400 $\mu m$ Cu allows homogenization of the Cu-0.2Cr (in wt\%) alloy. To achieve the maximum amount of Cr in Cu, in order to subsequently maximize the volume fraction of Cr precipitates during aging, the Cr layer thickness needs to be increased.  Additionally, simulations show that a 10 $\mu m$-thick Cr layer in contact with a 400 $\mu m$-thick Cu layer is too thick for the Cr to be fully incorporated in the Cu within 24~h and give a homogenized alloy. Improved Cu and Cr layer configuration is the subject of further investigation~\cite{Spathara:2025hrp}.

\subsection{Precipitation strengthening}
\begin{figure*}
    \centering
    \subfigure[\label{fig:2a}]{\includegraphics[width=0.333\linewidth]{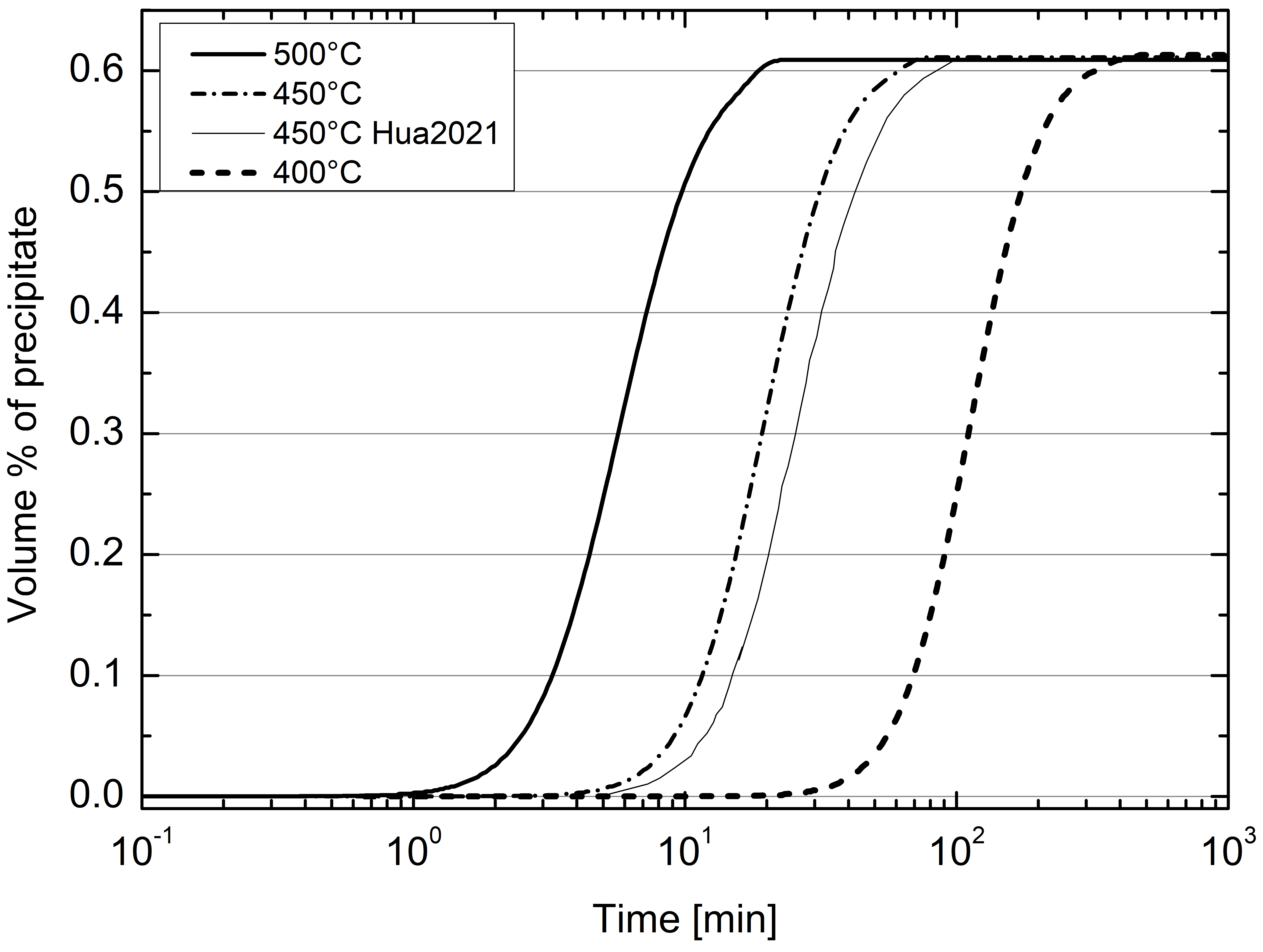}}
    \subfigure[\label{fig:2b}]{\includegraphics[width=0.333\linewidth]{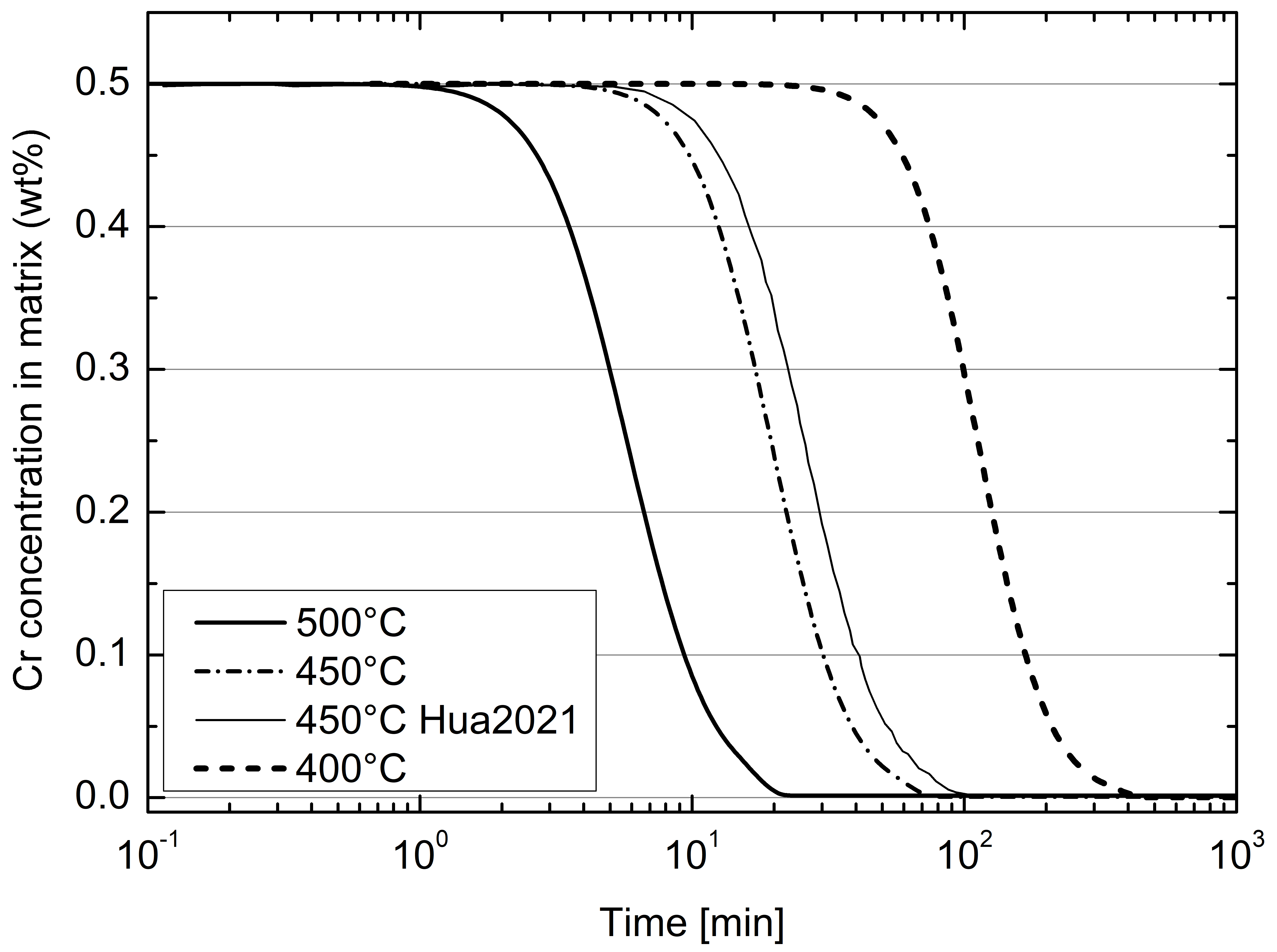}}
    \subfigure[\label{fig:2c}]{\includegraphics[width=0.32\linewidth]{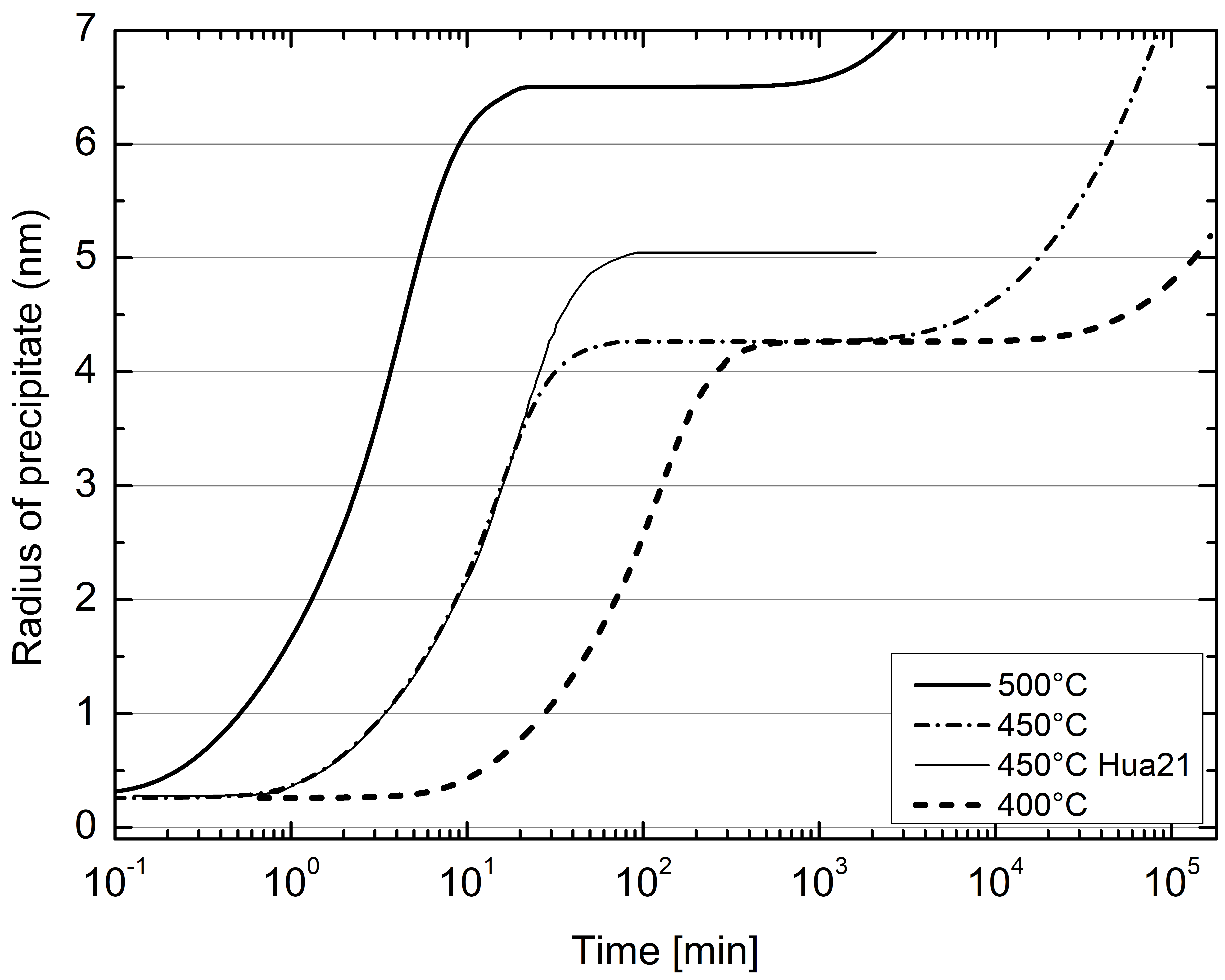}}
    \caption{Results of simulations of aging Cu-0.5Cr at 400$^\circ$C, 450$^\circ$C, and 500$^\circ$C, for \subref{fig:2a} the evolution of volume percentage of precipitate, \subref{fig:2b} Cr concentration in matrix, and \subref{fig:2c} mean radius. 
    Calculations (using Thermo-Calc Cu-based databases) corresponding to aging at 450$^\circ$C are included~\cite{HUANG2021102378}. 
    \label{fig:2}}
\vspace{-0.3cm}
\end{figure*}

Figure~\ref{fig:2} shows the simulated temporal evolution of the percentage of precipitate volume, Cr concentration in the matrix, and mean radius during aging of Cu-0.5Cr at 400$^\circ$C, 450$^\circ$C and 500$^\circ$C.

The temperature of 450$^\circ$C is selected as this is the main temperature for aging CuCr alloys in the literature, both in simulations and experiments of alloys manufactured from casting~\cite{HUANG2021102378}. The temperature of 500$^\circ$C is included in the simulations, as this is the temperature for aging in Ref.~\cite{Vitale:2021xrm}, where measurements are provided. The temperature of 400$^\circ$C is selected for comparison.

\begin{figure*}
    \centering
    \subfigure[\label{fig:3a}]{\includegraphics[width=0.325\linewidth]{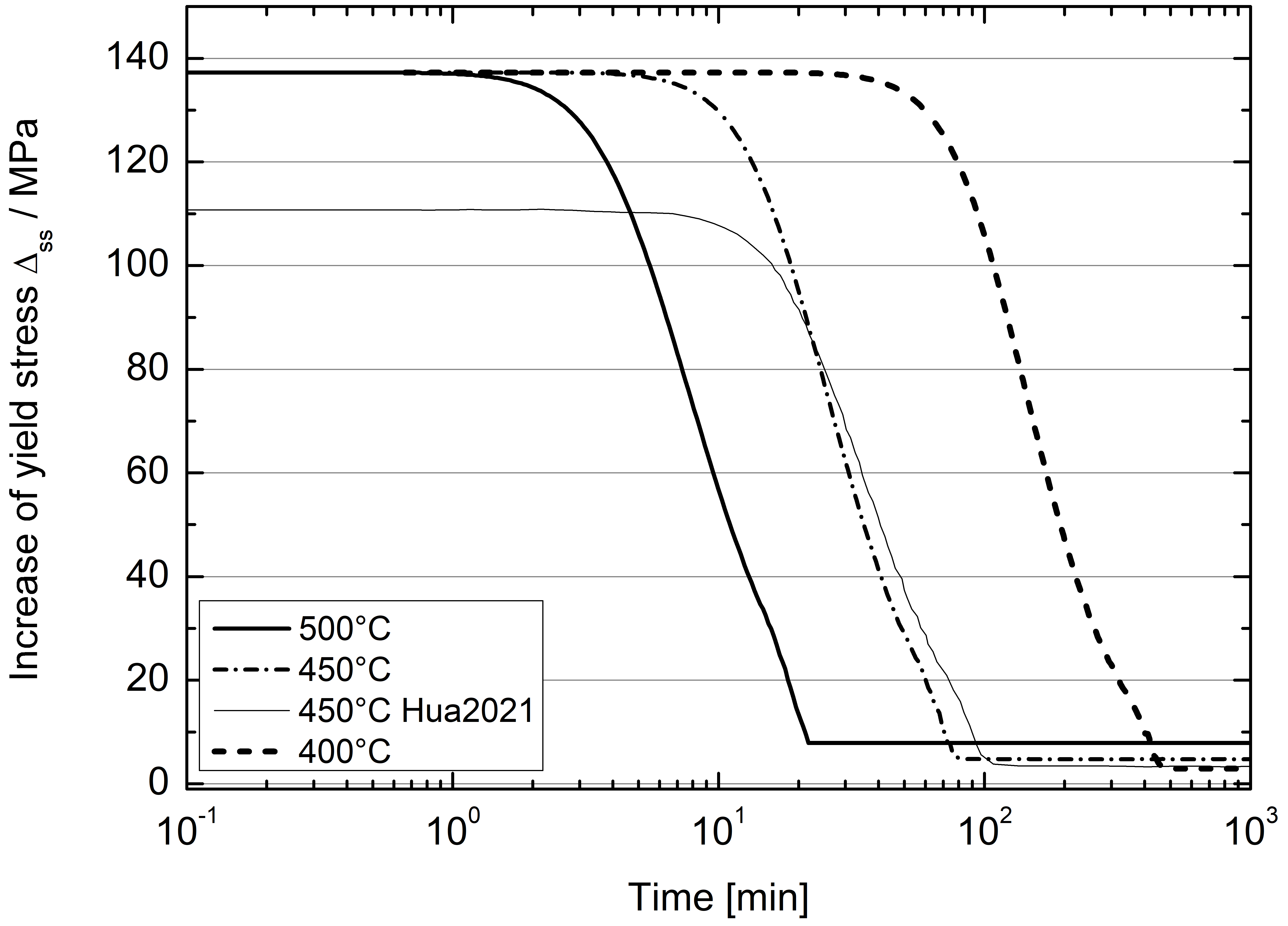}}
    \subfigure[\label{fig:3b}]{\includegraphics[width=0.325\linewidth]{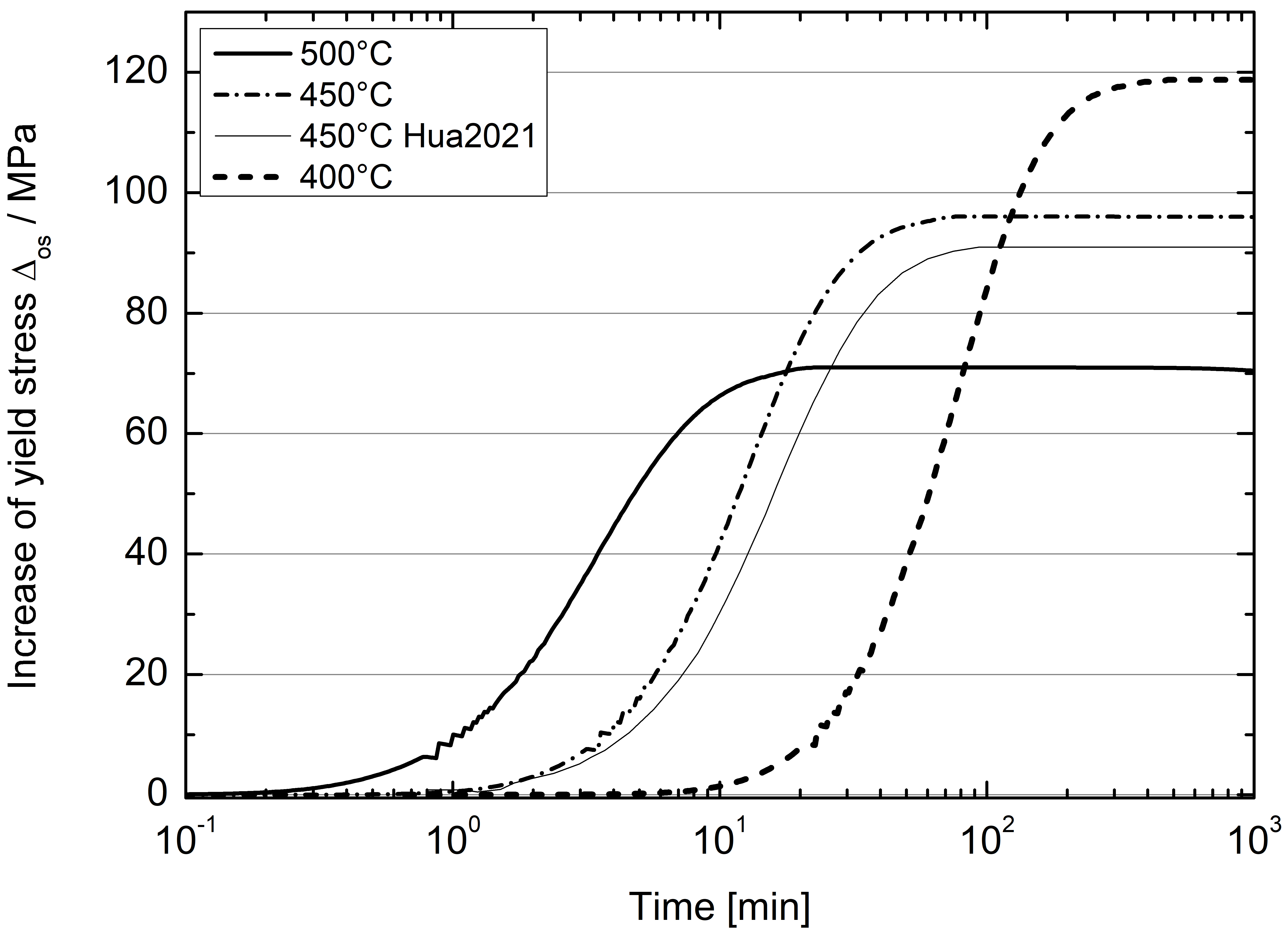}}
    \subfigure[\label{fig:3c}]{\includegraphics[width=0.325\linewidth]{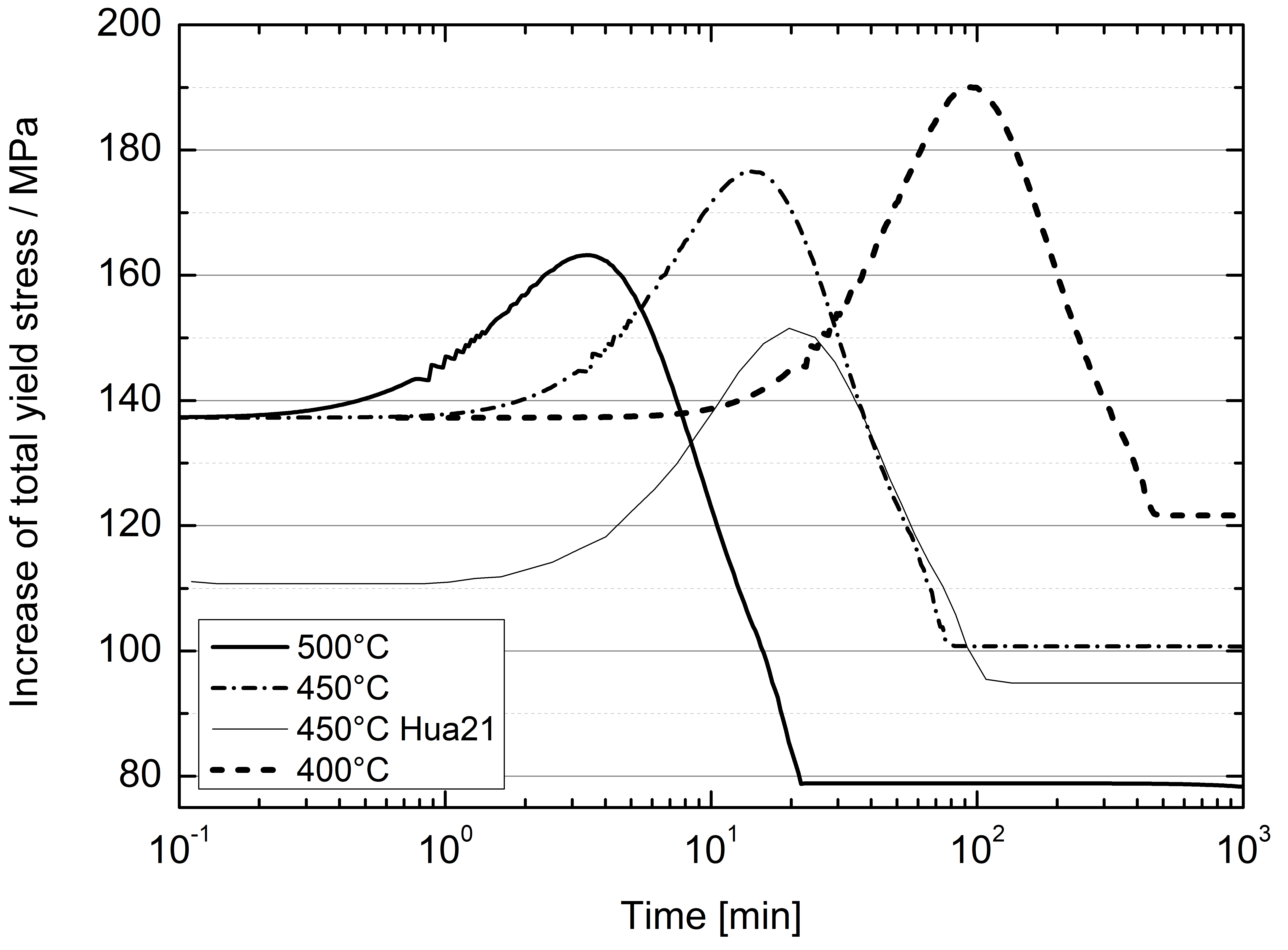}}
    \caption{Predictions for temporal evolution of yield strength due to \subref{fig:3a} solid solution stengthening, \subref{fig:3b} precipitation strengthening, and \subref{fig:3c} the sum of the two during aging at 400$^\circ$C, 450$^\circ$C, and 500$^\circ$C. The corresponding results for aging at 450$^\circ$C from Ref.~\cite{HUANG2021102378} are included. 
    \label{fig:3}}
    \vspace{-0.3cm}
\end{figure*}

Aging proceeds slower at lower temperatures. The percentage of volume fraction of the precipitating phase is limited to 0.61\% for all temperatures, corresponding to the maximum allowed for the specific alloy composition, reaching a plateau. Given that in this work, the precipitation strengthening of a CuCr alloy of 0.5~wt\% Cr content is explored, the volume fraction of the precipitating phase cannot exceed the particular value. The Cr concentration in the matrix decreases with time. It is 0.5~wt\% in the homogenized initial composition and as the precipitating phase bcc forms, Cr diffuses from the fcc matrix at almost 100\% of its initial concentration. The results of this work using TCHEA6 and MOBHEA3 are very similar to simulations from literature for the 450$^\circ$C~\cite{HUANG2021102378} using the Thermo-calc Software Cu-based alloys databases TCCU3 and MOBCU3~\cite{TCdatabase2, TCdatabase3}, especially in the \% of volume fraction of precipitates and the Cr concentration in the matrix. 

In terms of mean radius, the simulations of this work at 450$^\circ$C predict a smaller radius of precipitates compared to the simulation results reported in Ref.~\cite{HUANG2021102378}, which explains the higher yield strength due to precipitation strengthening predicted in this work, as shown in Fig.~\ref{fig:3b}. The final mean radius at 400$^\circ$C is predicted to be the same as that at 450$^\circ$C, although generally the maximum radius of the precipitates decreases with temperature. With smaller precipitates and the same volume fraction, even more enhanced mechanical properties are expected. 
This is shown in the graph with the prediction of the increase in yield strength due to precipitation strengthening (Fig.~\ref{fig:3b}). Interestingly, the same maximum radius on the plateau found for aging at 450$^\circ$C and 400$^\circ$C is not reflected in the yield strength due to precipitation strengthening. This can be explained by the fact that at lower temperatures precipitation growth proceeds slower and the volume fraction of the precipitating phase is still growing with time and has not yet reached its plateau, as shown in Fig.~\ref{fig:2a}. The comparison of the temporal evolution of the mean radius and the precipitate volume fraction becomes apparent in Eq.~\ref{eq10}.

Predictions for the evolution of the yield strength due to solid solution strengthening, precipitation strengthening, and the total yield strength can be found in Fig.~\ref{fig:3} during aging at 400$^\circ$C, 450$^\circ$C, 500$^\circ$C. The yield strength (or stress) due to solid solution strengthening data (Fig.~\ref{fig:3a}) is obtained by using Eq.~\ref{eq6} and due to the precipitation strengthening data (Fig.~\ref{fig:3b}) by using Eq.~\ref{eq9} and Eq.\ref{eq10}. 
Simulations have also revealed that the maximum total yield strength (or stress), that is, the sum of the corresponding data due to solid solution and precipitation strengthening, can be achieved by aging at a lower temperature for longer periods of time, as shown in Fig.~\ref{fig:3c}. 

At lower temperatures, Cr diffuses slower from the matrix toward the precipitates compared with aging at higher temperatures, and the volume fraction of Cr in the matrix is higher. This leads to an increase in the yield strength during both solid solution and precipitation strengthening. The relation is illustrated in the equations for the yield strength curves, Eq.\ref{eq6} and Eq.\ref{eq9}-\ref{eq10}, respectively. The yield strength evolution due to solid solution strengthening depends on the temporal Cr concentration in the matrix, so the trend of both is almost identical. The mole fraction of the initial concentration of Cr in the matrix for this work and in Ref.~\cite{HUANG2021102378} is identical to expected, corresponding to 0.5 wt\%~Cr. However, a much lower initial yield strength has been reported in Ref.~\cite{HUANG2021102378}, with the source of this difference remains unclear. 
A small deviation of the results of this work compared to Ref.~\cite{HUANG2021102378} is observed for the temporal evolution of the volume fraction of the precipitating phase, the Cr concentration in the matrix, and the size of mean radius for aging at 450$^\circ$C. This could be attributed to the different sets of CALPHAD-type thermodynamic and kinetic databases used in Ref.~\cite{HUANG2021102378}.

From this work, we deduce that when aging at 500$^\circ$C is performed, much shorter durations than those reported in Ref.~\cite{Vitale:2021xrm} are required to obtain optimal results for precipitation strengthening. Prolonged aging at 500$^\circ$C leads to an increase in the mean radius of the precipitates, as some precipitates grow larger and some others will shrink, indicating that coarsening is already taking place. Consequently, aging at lower temperatures should be further explored. 

\begin{figure}
    \centering
    \includegraphics[width=0.45\linewidth]{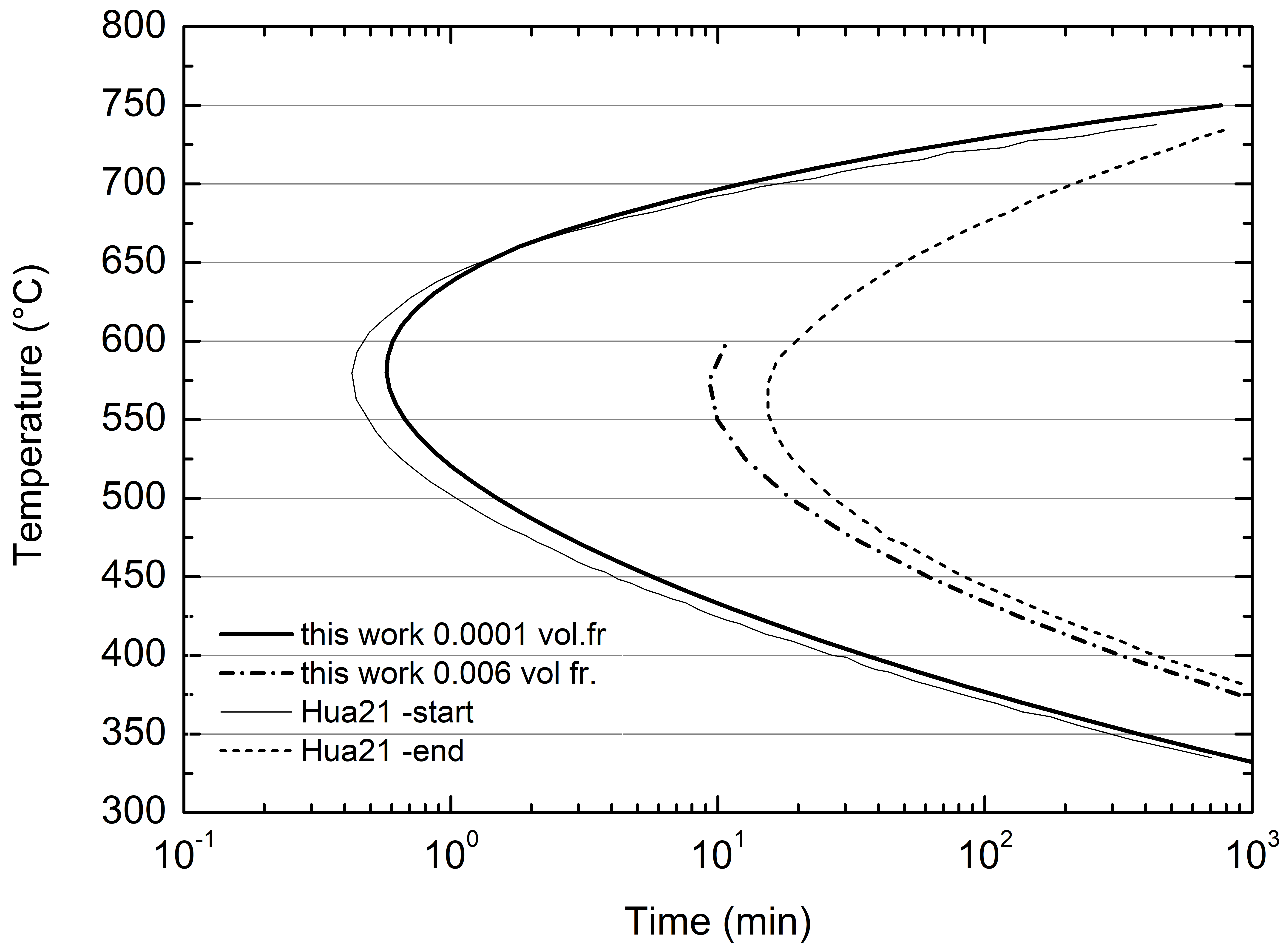}
    \caption{Calculation of TTT diagram for Cu-0.5Cr for 0.01 and 0.6 volume fraction \% using TCHEA6 and MOBHEA3. 
    Calculations (using Thermo-Calc Cu-based databases) are included~\cite{HUANG2021102378}. 
    \label{fig:4}}
    \vspace{-0.3cm}
\end{figure}

Figure~\ref{fig:4} shows the Time-Temperature-Transformation (TTT) diagram calculated using the TCHEA6 and MOBHEA3 databases. 
This diagram is useful to understand the time-temperature dependence of growing a new phase, starting from a specific microstructure. According to this TTT diagram the maximum percentage of the volume fraction of 0.61\% can be reached after 10-12 minutes, 60 minutes and 300 minutes for aging  at 500$^\circ$C, 450$^\circ$C and 400$^\circ$C, respectively. 

Moreover, the diagram is compared to the corresponding diagram in Ref.~\cite{HUANG2021102378}.
The set of High Entropy alloys databases (TCHEA6/ MOBHEA3) used in this work predict a narrower region at which bcc precipitation can occur in a wide range of aging temperatures. The observed differences are attributed to the different databases used. 
The calculations in Ref.~\cite{HUANG2021102378} were performed using the Cu-based alloys databases TCCU3 and MOBCU3. These latter calculations are validated with measurements, but not yet for radiopure alloys. 

From the above comparisons, it is seen that any differences in the simulation results when different sets of databases are used are typically small.
Further validation of databases will determine their suitability or may lead to further work developing more accurate descriptions through dedicated databases for radiopure Cu-based alloys. Already, this work validates the High Entropy alloys database set for the stage of solution heat treatment, through the presented comparisons with measurements reported in Ref.~\cite{Vitale:2021xrm}. In the prediction of the temporal evolution of yield strength due to solid solution and precipitation strengthening, further comparisons with the experimental data would be beneficial. For example, the value of interfacial energy between the bcc and fcc phases is crucial, especially when the nucleation barrier is determined as shown in Eq.\ref{eq5}. Due to the lack of an experimental value for the interfacial energy, this was taken from calculations reported in Ref.~\cite{HUANG2021102378} using the TCCU3 and MOBCU3 databases and may affect predictions of the evolution of yield strength. This is not very easy to obtain experimentally. Further experimental measurements in yield strength will be beneficial for more accurate predictions.

Overall, the methodology presented in this work using the Thermo-calc modlules DICTRA and TC-PRISMA is evidently a powerful approach to develop predictive tools to design and optimize radiopure alloys and their thermal processing. 

\section{Conclusion}
Thermally processed, electroformed copper-chromium (CuCr) alloys are stronger than EFCu and have the potential to maintain high radiopurity. As a result, they are crucial for the development of next-generation rare-event search experiments.

The processes of solution heat treatment and aging of CuCr alloys have been modeled using computational thermodynamics. The simulation results are compared with the measurements available in the literature.

Furthermore, a methodology is proposed to predict materials properties and optimize manufacturing processing, using DICTRA and TC-PRISMA simulations. This can lead to the acceleration of the design of the Cu and Cr layer configuration in the fabrication stage and the optimization of the parameters for thermal processing to achieve the maximum enhancement of mechanical properties.

Future work includes optimizing the Cu and Cr layer configuration to achieve homogeneous alloy compositions with up to 0.5~wt\% Cr during solution heat treatment, and of the aging process aiming at lower temperatures and durations while further enhancing the alloy strength. These need to be achieved without compromising the radiopurity and electrical conductivity requirements.

Manufacturing the radiopure CuCr alloys designed using the methodology proposed in this work and their characterization will further determine the suitability of the thermodynamic and kinetic databases used and will contribute to the development of dedicated databases for this application.

\section{Declaration of competing interest}
The author declares that they have no known competing interests or personal relationships that could appear to influence the reported work.

\section{Acknowledgement - Funding}
Support from the UKRI Horizon Europe Underwriting scheme (GA101066657/Je-S EP/X022773/1) is acknowledged. 
The author acknowledges also the valuable discussions with Prof. Konstantinos Nikolopoulos, Mr. Eric Hoppe and Dr. Patrick Knights. 

\bibliographystyle{elsarticle-num}
\bibliography{bibliography}
\end{document}